\documentclass[a4paper,12pt]{article}
\usepackage{amsfonts}
\usepackage{lscape}
\begin{document}

\makeatletter
\@addtoreset{equation}{section}
\makeatother
\renewcommand{\theequation}{\thesection.\arabic{equation}}

\def\tr#1{\mbox{tr}\,#1}

\def\cosx#1{{\tt c}{\left(#1\right)}}
\def\coshx#1{{\tt K}{\left(#1\right)}}
\def\sinhx#1{{\tt S}{\left(#1\right)}}
\def\psip#1{\psi^+_{#1}}
\def\psim#1{\psi^-_{#1}}
\def\LambdaTensor{{\tt LambdaTensor1.0}}

\thispagestyle{empty}
\begin{flushright}
AEI-2002-065
\end{flushright}

\begin{center}
{\bf\Large Introducing \LambdaTensor{} -- A package for explicit symbolic and numeric Lie algebra and Lie group calculations}

\bigskip\bigskip\bigskip

{\bf T. Fischbacher\medskip\\ }
{\em Max-Planck-Institut f\"ur Gravitationsphysik,\\
     Albert-Einstein-Institut,\\
     M\"uhlenberg 1, D-14476 Potsdam, Germany\\ }
\smallskip
{\small \tt tf@aei-potsdam.mpg.de}\bigskip

\end{center}

\begin{abstract}
Due to the occurrence of large exceptional Lie groups in supergravity,
calculations involving explicit Lie algebra and Lie group element
manipulations easily become very complicated and hence also
error-prone if done by hand. Research on the extremal structure of
maximal gauged supergravity theories in various dimensions sparked
the development of a library for efficient abstract multilinear algebra
calculations involving sparse and non-sparse higher-rank tensors, which is
presented here.
\end{abstract}

\vfill
\leftline{{August 2002}}

\newpage

\section{Introduction}

Although nowadays, there are many powerful general-purpose symbolic
algebra packages available, the most well-known ones probably being
{\tt Mathe\-mati\-ca}, {\tt Maple}, {\tt MuPAD}, {\tt MACSYMA}, and {\tt
Reduce}, there still are many open problems which require (sometimes
highly specialized) computer algebra that has become feasible on
modern machines, but is beyond the scope and limitations of these
systems. Hence, in some areas of research, there is notable demand for
tailored symbolic algebra.

In this work, a package for group-theoretic calculations is presented
which originally was created for the computation of explicit
expressions for the potentials of gauged maximal supergravity theories
on Euler-angle-like para\-me\-trized submanifolds of the manifolds of
supergravity scalars. Currently, this seems to be the most effective
tool for the determination of these potentials and hence possible
vacua of these supergravity theories, but it may be applicable to a
wider class of more general group-theoretic problems. Nevertheless,
the structure of mentioned supergravity potentials turned out to be so
rich that they have not been mined completely up to date, which by
itself is a strong incentive to make this package publicly available.

There are two complementary ways how such a package could be
implemented: either as a library manipulating tensor expressions on
the symbolic level, like {\tt MathTensor}, or as a library operating on
explicit tensor coordinates (which nevertheless may well be symbolic
themselves). Here, the latter approach was chosen for two reasons:
first, the possibility to always go down to plain numbers on the fly
leads to better testability and broader applicability in the field of
numerical calculations, and second, since one typical application is
to break the exceptional group $E_{8(+8)}$ to subgroups as small as
possible, the number of terms generated on the purely symbolic level
would explode, nullifying the conceptual advantages of the former
approach.

Some of the outstanding features of the symbolic algebra framework
presented here are:
\begin{itemize}

\item It is Free Software, available under the GNU Lesser General Public 
License, Version 2.1. (Nevertheless, users are urged to read the
accompanying copyright information to see how it is applied in this
particular case.)

\item It is complementary to the {\tt LiE} package in that sense that {\tt LiE}
can not perform calculations that deal with explicit Lie algebra or
Lie group elements, while such functionality lies at the heart of
\LambdaTensor.

\vfill

\item It is reasonably efficient, since it employs and is designed for
an optimizing machine code compiler.

\item It is extensible {\em at the level of the implementation itself}, not,
like most other symbolic algebra packages, only at the level of a
built-in application language. In particular, {\em it does not restrict
the user of this package to a simple stripped-down C-like-in-spirit
application language}. Instead, the user has full access to the
feature-rich, extensible, compiler system on which this package was
built, CMU Common LISP.

\end{itemize}

It is perhaps noteworthy that utilizing a functional subsystem (in
this case, Common LISP) not only obsoletes the need for the
implementation of yet another application-centric programming
language, which hardly could and should compete with a system backed
by its own user community and team of developers, but also readily
makes available a variety of other useful existing libraries and tools
written for this system. The rationale behind the design decision to
use Common LISP, and in particular to build on the CMU CL
implementation, instead of using one of the other viable alternatives
suggesting themselves here, namely Haskell (in particular, GHC),
Objective Caml, or Scheme (in particular, Gambit), which is of course
mainly a question of personal preferences of the author, comes in part
from the experienced inflexibility of systems based on typed
$\lambda$-calculus (ML and Haskell -- although the power of Haskell's
typeclass system has to be acknowledged), partly due to general
typability problems, partly since these systems do not provide means
to modify the very language itself which would be comparable to the
macro facilities of LISP (although this is much less a problem in a
lazy functional language like Haskell), and in part from the baroque
richness in features of the Common LISP standard (especially when
compared to Scheme). The main drawback of this decision is that the
most advanced freely available Common LISP compiler system, CMU CL, is
only well supported for x86-based platforms, and due to technical
reasons, comes with a limitation on the size of the data being
processed in memory of 800 MB\footnote{Experienced Unix users can
raise this to about 1.6 GB, but in many cases this requires in
particular manual modification of the kernel source.}.

The central idea behind this package is to utilize the observation
that tensors showing up in group theory calculations frequently are
very sparsely occupied -- for example, using the conventions of
$\cite{Fischbacher:2002fx}$, structure constants of the largest (248-dimensional)
exceptional Lie group $E_8$ $f_{\mathcal{A}\mathcal{B}}{}^\mathcal{C}$
contain only $49\,440$ out of $248^3=15\,252\,992$ nonzero entries --
and hence, we can make good use of efficient implementations of
abstract algorithms that can handle sparsely occupied higher-rank
tensors. Efficient code working on sparse matrices is widely used and
readily available; the appropriate algorithms for handling higher-rank
tensors are also quite well-known, albeit in a very different context:
relational databases.

In particular, at the level of explicit tensor entries, forming a
quantity like
\begin{equation}
        M_{abc}=N_{gha}P^{gh}{}_{bc}
\end{equation}
translates as follows into the language of relational databases (SQL syntax used here):
\begin{verbatim}
SELECT t1.index3 as index1,
       t2.index3 as index2,
       t2.index4 as index3,
       SUM(t1.val*t2.val)
   FROM tensor1 t1, tensor2 t2
   WHERE t1.index1=t2.index1 AND t1.index2=t2.index2
   GROUP BY t1.index3, t2.index3, t2.index4;
\end{verbatim}

Unfortunately, it is not feasible to just connect to an existing SQL
data\-base system (like PostgreSQL), create relations for tensors, and
use existing implementations of these algorithms by doing all the
calculations in the database, for various reasons. Besides
considerations concerning the efficiency of communication, and
considerable additional computational overhead due to data\-bases having
different aims, one major problem is that extending the database
system to abstract from the implementation of sum and product here, as
is necessary as soon as we want to work with data types not natively
supported by the database (which are frequently limited to integers
and floatingpoint numbers) would bring along too many technical
problems. Hence, what is required is a re-implementation of the
underlying database algorithms with numerical and symbolic tensor
computations as applications in mind. Furthermore, this implementation
has to be abstract enough to allow all relevant arithmetic operations
to be provided as parameters, so that one may switch between
approximate numerics, exact (i.e. rational number) numerics, and
symbolic calculations. (The ability to implement and use arbitrary
arithmetics on tensor entries has proven to be of great value during
the debugging phase of the symbolic algebras provided within this
package. For example, it is easy to lift an existing implementation of
arithmetic operations on symbolic terms to an implementation working
on pairs of terms and numerical values of these terms for a given
occupation of variables that signals an error whenever a discrepancy
between these values shows up.)

\vfill\newpage

The \LambdaTensor{} package consists of the following parts:
\begin{enumerate}

\item Various general-purpose functions providing important infrastructure.
(containing e.g. simple combinatorial functions, basic linear algebra,
balanced binary trees, simple optimization functions, priority queues,
and a serializer.) For the Debian GNU/Linux system, this part is
available as a separate package.

\item Support for sparse higher-rank arrays and tensor operations on them,
where implementations of arithmetic operations on tensor entries may
be given as parameters.

\item Different implementations of symbolic algebra.
(One which is similar in spirit and intention, though not in scope, to
conventional general-purpose symbolic algebra packages, one which is
aggressively optimized (and hence far from being general-purpose) for
calculations involving products of trigonometric functions of the
particular form showing up in supergravity calculations as in
\cite{Fischbacher:2002hg,Fischbacher:2002fx}, and a third one utilizing the CMUCL port of the MAXIMA
symbolic algebra package.)

\item Applications. In particular, definitions relevant
for the exceptional groups $E_{7(+7)}$ and $E_{8(+8)}$ and the
potentials of maximal gauged extended supergravity theories in three
and four dimensions.

\item Worked out, documented examples that demonstrate how to use the package.

\end{enumerate}

Since this package was created as a byproduct of work targeted at the
determination of the extremal structure of supergravity potentials,
these tools are in some aspects just as good as they had to be for
this task, with lots of opportunities for optimization and improvement
still remaining. In a different vein, since those particular
calculations are quite demanding, these tools {\em are} quite
optimized in the most central aspects. Nevertheless, large parts of
this codebase are constantly exchanged, improved, re-written, and
hence, major changes should be expected between version 1.0 and
subsequent versions.

\section{An overview over \LambdaTensor}

Since detailed technical documentation can be found in the
\LambdaTensor{} manual, we only want to give a brief overview over
concepts and algorithms underlying the different pieces of this
package.

\subsection{General purpose functions}

This is a collection of various functions and macros ranging from
simple more convenient redefinitions of features already available in
COMMON LISP to implementations of ubiquitous algorithms to complex
facilities providing vital infrastructure for the other parts of
\LambdaTensor. Since parts of this highly inhomogeneous conglomerate
of functions and definitions are not essential for \LambdaTensor, but
perhaps useful in a much broader context, the decision was made to
split this off into a separate package for the the Debian distribution
of \LambdaTensor. Current functionality provided here encompasses, but
is not limited to, macros providing machine-code optimization
information to the compiler, various abbreviations and definitions
that were inspired by the Perl language, elementary combinatorial
functions, basic polynome factorization, linear algebra, and
optimization support, as well as efficient implementations of balanced
binary trees and priority queues.

\subsection{Sparse array functions}

This is the heart of \LambdaTensor, implementing sparse arbitrary-rank
arrays using database algorithms. In particular, a sparse array is
represented internally as a multidimensional hash of its nonzero
components. Sparse arrays are transparently re-hashed if their
occupation density grows, up to a certain percentage, where the
implementation internally switches to storing tensor entries in a
nonsparse array. Currently, removing entries from a sparse-array does
not induce the underlying hash to shrink once occupation density falls
below a certain level.

The most important sparse array function provided is {\tt SP-X} which
implements efficient tensor multiplication, contraction, and index
reordering of an arbitrary number of tensors (limited by resources)
where multiplications and contractions are heuristically sequenced in
such a way to minimize the total number of operations. To give an
example of what {\tt SP-X} can do and how it is used, let's assume
that the variable {\tt F-abc} contains the structure constants
$f_{ab}{}^c$ of a Lie algebra. Then, the Cartan-Killing metric
$g_{ab}=f_{ap}{}^q\,f_{bq}{}^p$ can be computed as follows:
{\small
\begin{verbatim}
(defvar metric (sp-x `(a b) `(,F-abc a p q) `(,F-abc b q p)))
\end{verbatim}
}
A second example: assuming {\tt so8-sigma} is the
rank-3 tensor of $SO(8)$ $\Gamma$-matrices with index order
$(i,\alpha,\dot\beta)$, the following piece of code checks
the Clifford algebra properties under contraction of the
cospinor indices:

\vfill
{\small
\begin{verbatim}
(sp-multiple-p
 (sp+ (sp-x `(i j a b) `(,so8-sigma i a a*) `(,so8-sigma j b a*))
      (sp-x `(i j a b) `(,so8-sigma j a a*) `(,so8-sigma i b a*)))
 (sp-x `(i j a b) `(,(sp-id 8) i j) `(,(sp-id 8) a b)))
\end{verbatim}
}

Besides tensor arithmetics (parametrized by the underlying
implementation of arithmetics on tensor entries)\footnote{for some
kinds of tensor entries, in particular complex double-precision
floatingpoint numbers, the implementation uses specially optimized
versions of internal functions.} and functions computing embedding
tensors for index split operations, this piece of code also provides
linear algebra functions on sparse tensors (like {\tt SP-INVERT} and
{\tt SP-MATRIX-EIGENVALUES} for quadratic rank-2 tensors over numbers
(not yet symbolic expressions)), conversion functions mapping sparse
arrays to and from nonsparse vectors, as well as group theory related
functions like {\tt SP-LIN-INDEP-COMMUTATORS}, giving a linearly
independent basis for all the commutators of two sets of quadratic
sparse rank-2 tensors, or {\tt SP-FIND-ROOT-OPERATOR} which, given
structure constants, a Cartan subalgebra, and a root vector,
determines the adjoint-representation coefficients of the
corresponding ladder operator.

\subsection{Symbolic Algebra}

There are different implementations of symbolic algebra available
within this package, each of them having its own raison d'\^etre. The
most effective since most highly optimized towards the problem for the
original task of computation of supergravity potentials is the {{\em
packof-exp}, or, in brief, {\em poexp}} algebra. When looking at a
typical supergravity potential restricted to a gauge subgroup singlet
manifold, like the following one from
\cite{Fischbacher:2002fx} of $N=16, D=3$ supergravity with gauge group $SO(8)\times SO(8)$
on the conveniently parametrized four-dimensional $\left(SL(2)/U(1)\right)^2$
manifold of $G_{2,{\rm diag}}$ singlets,
{\small
\begin{equation}
\begin{array}{lcl}
-8g^{-2}V&=&\frac{243}{8}+\frac{7}{2}\,\cosh(2\,s)+\frac{49}{8}\,\cosh(4\,s)+\frac{1141}{64}\,\cosh(s)\,\cosh(z)\\
&&+\frac{427}{64}\,\cosh(3\,s)\,\cosh(z)-\frac{7}{64}\,\cosh(5\,s)\,\cosh(z)\\
&&-\frac{25}{64}\,\cosh(7\,s)\,\cosh(z)+\frac{21}{8}\,\cos(4\,v)\\
&&-\frac{7}{2}\,\cos(4\,v)\,\cosh(2\,s)+\frac{7}{8}\,\cos(4\,v)\,\cosh(4\,s)\\
&&-\frac{21}{64}\,\cos(4\,v)\,\cosh(s)\,\cosh(z)\\
&&+\frac{21}{64}\,\cos(4\,v)\,\cosh(3\,s)\,\cosh(z)\\
&&+\frac{7}{64}\,\cos(4\,v)\,\cosh(5\,s)\,\cosh(z)\\
&&-\frac{7}{64}\,\cos(4\,v)\,\cosh(7\,s)\,\cosh(z)\\
&&-\frac{1645}{128}\,\cos(v-w)\,\sinh(z)\,\sinh(s)\\
&&+\frac{651}{128}\,\cos(v-w)\,\sinh(z)\,\sinh(3\,s)\\
&&+\frac{7}{128}\,\cos(v-w)\,\sinh(z)\,\sinh(5\,s)\\
&&-\frac{49}{128}\,\cos(v-w)\,\sinh(z)\,\sinh(7\,s)\\
&&-\frac{315}{64}\,\cos(3\,v+w)\,\sinh(z)\,\sinh(s)\\
&&+\frac{133}{64}\,\cos(3\,v+w)\,\sinh(z)\,\sinh(3\,s)\\
&&-\frac{7}{64}\,\cos(3\,v+w)\,\sinh(z)\,\sinh(5\,s)\\
&&-\frac{7}{64}\,\cos(3\,v+w)\,\sinh(z)\,\sinh(7\,s)\\
&&+\frac{35}{128}\,\cos(7\,v+w)\,\sinh(z)\,\sinh(s)\\
&&-\frac{21}{128}\,\cos(7\,v+w)\,\sinh(z)\,\sinh(3\,s)\\
&&+\frac{7}{128}\,\cos(7\,v+w)\,\sinh(z)\,\sinh(5\,s)\\
&&-\frac{1}{128}\,\cos(7\,v+w)\,\sinh(z)\,\sinh(7\,s),
\end{array}
\end{equation}
}
one notices that the typical summand in such a term (as well as in all
intermediate quantities) is of the form $k\cdot\exp\left(\sum_j c_j
v_j\right)$, where $v_j$ are variable names, and $c_j$ as well as $k$
all are either real or imaginary rational numbers. Furthermore, these
terms `come in packs' and can be grouped together to form summands
like $\frac{7}{128}\,\cos(7\,v+w)\,\sinh(z)\,\sinh(5\,s)$. Hence, we
can forge an internal representation of such terms which is orders of
magnitude more efficient both in terms of memory consumption and
possible reductions than the conventional one of a generic term (as
used by other computer algebra systems) by using this additional
structure. (Details are given in the manual.)

It is the combination of this problem-specific implementation of a
symbolic algebra with efficient sparse array database algorithms that
made it possible to transcend all previous limitations in complexity
in \cite{Fischbacher:2002fx}.

As is perhaps imaginable, this aggressively optimized symbolic algebra
was not the first one to be employed in conjunction with sparse tensor
algorithms. The former one, which is available as the
function-polynome ({\em funpoly}, or often briefly {\em fp}) algebra,
is much closer in design to conventional symbolic algebra, hence also
more flexible and still shows up in some places within
\LambdaTensor. (Note that the {\em poexp} algebra is so specialized
that it can not handle anything else but terms of the structure
described above. In particular, it can not represent quasipolynomials,
hence one easily runs into trouble when nilpotence enters the stage.)
The funpoly symbolic algebra also implements some peculiarities going
beyond what one may expect from conventional symbolic algebra
packages, in particular some non-local reductions of precisely that
kind which FORM avoids by construction in order to be able to
efficiently handle formulae much bigger than available memory. Since
this piece of code did not undergo as many evolutionary cycles of
being re-written as some of the rest of this package, its design still
shows some flaws and weaknesses\footnote{in particular, handling of
fractional powers of rational numbers is quite clumsy}, and so it is
scheduled for replacement in later versions.

Since it may nevertheless be important to also have a flexible,
general, tested, and powerful implementation of some symbolic algebra
available that can be used in conjunction with this package, even if
one should not try to use that particular one for the calculation of
supergravity potentials, \LambdaTensor{} also comes with a simple
interface to the free {\tt MACSYMA}-replacement {\tt MAXIMA}, which
was originally implemented on top of {\tt GCL}, but then also ported
to other LISPs, including CMU CL.

\subsection{Applications and examples}

\LambdaTensor{} comes with optional additional definitions related to the
groups $E_{8(8)}$, $E_{7(7)}$ as well as important subgroups thereof
and further functions relevant for the computation and investigation
of the structure of the scalar potentials of three- and
four-dimensional maximal gauged supergravity theories.
Finally, detailed worked-out examples are provided within the package
which explain how to apply it to supergravity calculations.

\section{Availability and concluding remarks}

The most recent version of {\tt LambdaTensor} is available from the webpage
{\tt http://www.cip.physik.uni-muenchen.de/\~\relax tf/lambdatensor/}

To the author's best knowledge, the library presented here is the
first abstract implementation of efficient fundamental sparse
higher-rank tensor multilinear algebra, thus possibly closing an
important gap. Therefore, the present author considers a release under
a free software license, in particular, the GNU LGPL 2.1, as
adequate. As stated in detail in the accompanying copyright
information, users of this library are asked to quote the present
article, since it is customary to use citations as a rough measure for
scientific relevance, so that further development of this work can be
kept up to the needs of its users.

\section{Addendum: New features in version 1.1}

Since an update of this software package hardly justifies a new paper,
a brief overview over new functionality introduced in version 1.1,
released on 25.03.2003, is given in this addendum. First and perhaps
most important, the installation process has been considerably
simplified, now providing packages for all major Linux distributions.
Besides some minor bugfixes to code and documentation, a considerable
amount of new functionality has been implemented to extend the
group-theoretic capabilities of this library into the direction of the
LiE program.  Among the major new algorithm implementations are a
version of the Fast Freudenthal algorithm to calculate weight
multiplicities of representations of simple groups, the Peterson
recursion formula to determine root multiplicities of Kac-Moody
algebras (this was used in \cite{Nicolai:2003fw} to calculate level
decompositions of the infinite-dimensional algebras $E_{10}$ and
$E_{11}$), as well as LISP-oriented versions of the Todd-Coxeter coset
enumeration algorithm and the Schreier-Sims algorithm for permutation
groups.

% \newpage

\begingroup\raggedright\endgroup

\end{document}